\documentclass{iaus}
\usepackage{graphicx}

\title[Gaia SOC] 
{Gaia Science Operations Centre}

\author[W. O'Mullane \& U. Lammers]   
{William O'Mullane$^1$, Uwe Lammers$^1$, John Hoar$^1$, Jose Hernandez $^2$ }

\affiliation{
$^1$European Space Astronomy Centre, Avda. de los Castillos s/n
E-28692,  Villanueva de la Canada (Madrid)  \\[\affilskip]
$^2$Aurora based at ESAC \\[\affilskip]
}

\pubyear{2008}
\volume{xxx}  
\pagerange{119--126}
\date{\today and in revised form ??}
\jname{Proceedings Title IAU Symposium}
\editors{A.C. Editor, B.D. Editor \& C.E. Editor, eds.}
\begin{document}

\maketitle



\firstsection 
\section{Introduction}
The SOC hosts the data hub for the Gaia data processing (Sect. \ref{sect:mdb}) and must carefully
 balance ESA style standards (and bureaucracy) with the scientific communities creativity and dedication. 
 SOC is closely 
involved with the DPAC (Sect \ref{sect:dpac}) and is the communities interface to the satellite through MOC.
 SOC takes a  direct development role not just in the operations software such as POS (Sect. \ref{sect:pos})
  but in some of the science processing software such as the Astrometric Global Iterative Solution (AGIS) 
  \cite{Hobbs08}.  SOC is a major contributor to the GaiaTools library (Sect. \ref{sect:gaiatools}). During mission 
  operations SOC will run both operations type software as well as science processing systems such as IDT (Sect \ref{sect:idt})
  and AGIS.
  Some further details are given bellow.

\section{Data Processing and Analysis Consortium (DPAC)} \label{sect:dpac}
In May 2007 with approval from ESA's Science Programme Committee (SPC) 
DPAC was formalized as a pan-European consortium in charge of designing, implementing, and operating the 
data processing system needed to create the Gaia catalogue.
DPAC adopted the ECSS (European Cooperation on Space Standardization) software engineering standards and
practices tailored for the Gaia project while following 
a cyclic development model in which software modules are incrementally built. The time to launch is
subdivided into phases of six months durations. A large emphasis is placed on 
unit/integration and end-to-end system testing. 

\section{Payload Operations System (POS)} \label{sect:pos}
The function of the POS is to support all payload
operations activities performed at the Gaia Science Operations Centre (SOC).
The POS shall support the following activities:
\begin{itemize}
\item Monitoring the execution of the Nominal Scanning Law
\item Calculation of transition times between the Nominal and Modified Scanning Law
\item Generation of the Science Schedule containing the predicted on-board
data rate according to the Scanning Law, galaxy and instrument models
\item Tracking the current status and history of configuration parameters of
payload
\item Creating and dispatching of Scientific Operations Requests, including the
generation of updates to configuration parameters of payload systems
\item Tracking the execution of Scientific Operations Requests through the
Mission Timeline and Telecommand History
\item Processing of optical observations of Gaia received from DPAC into the
required format for the MOC (for improving Gaia position and velocity
data from standard ranging measurements)
\end{itemize}


\section{Gaia Main Database (MDB)} \label{sect:mdb}
The MDB is the central repository for all the
data produced by Gaia and the DPAC. The system shall be operated
at ESAC but its development and definition involves all of the
the CUs.
The MDB is the hub for all Gaia processing  in
that it provides the input data for, and receives the
generated output from, all processing systems. The details of these relations are specified in the
MDB Interface Control Document (ICD).
The MDB will hold a range of different data types  and 
the total data volume that the MDB will ultimately contain is still
somewhat uncertain but it shall be of the order of hundreds of
Terabytes and could even reach the Petabyte range.

The MDB will contain a number of version-controlled results data
bases that will form a sequence $V_0$, $V_1$ \ldots $V_n$. Each $V_m$ comprises
three main parts, viz. {\em raw data}, coming from the satellite, {\em intermediate data} 
generated by the pre-processing pipeline and {\em reduced data}
, coming from the various processing systems. The reduced data in any $V_m$ is only produced from
data contained in the previous $V_{m-1}$. The versions will be produced cyclically  
at regular intervals, e.g. every 6 months.

\section{GaiaTools library  and Parameter DB} \label{sect:gaiatools}
GaiaTools is a common software library collaboratively written in Java to support and
facilitate the software development in DPAC. The principle reasons for GaiaTools are to 
avoid duplication of effort and reduce errors in common routines.
Policy decisions are taken by a GaiaTools committee composed of one representative from
each CU
All code development is done by DPAC members in a distributed
manner following DPAC's software engineering guidelines.
Functionality in GaiaTools range from the Data Access Layer (DAL) and plotting 
through general numerical routines (Vector+Matrix algebra, interpolation etc) to
Solar System ephemeris access.

The Gaia Parameter Database (GPDB) is a central, searchable
repository of all mathematical, physical, mission, satellite, and payload
design parameters with relevance for Gaia. A Java export of the GPDB
is distributed together with the GaiaTools library.

\section{Initial Data Treatment } \label{sect:idt}
IDT can be regarded as a pre-processing pipeline
converting raw telemetry from the satellite into higher level
data products for downstream software (like
AGIS ).
It will be operated in a quasi-continuous manner at SOC to
process telemetry in near-real time. Telemetry data
will go into a {\em Raw Database} - processed data into a
IDT/FL database and from there into the Main Data
Base (Sect. \ref{sect:mdb}.
The primary IDT outputs will be:
Extracted images parameters (centroids, fluxes),
Match-table linking observations to sources,
and Refined on-board attitude.

The system is developed by DPAC within
CU3  with major contributions
from University of Barcelona, Spain.

%

\section{Conclusions}\label{sec:concl}

The SOC has a diverse role to play in the Gaia data processing in conjunction with
and as part of the DPAC and look forward to meeting the Gaia processing challenges.

\end{document}